\documentclass[conference]{IEEEtran}
\IEEEoverridecommandlockouts

\usepackage[dvipsnames]{xcolor}
\usepackage[utf8]{inputenc} 
\usepackage{comment}
\usepackage[T1]{fontenc}
\usepackage{url}
\usepackage{ifthen}
\usepackage{cite}
\usepackage{amsmath}
\usepackage{mathtools,cuted}
\usepackage{amssymb} 
\usepackage{cite}
\usepackage{pgf,tikz,pgfplots}
\usepackage{algorithmic}
\usepackage{graphicx}
\usepackage{tabularx}
\usepackage{multirow}
\usepackage{array}
\usepackage{tikz}
\usepackage{ctable}
\usepackage{stfloats}

\newcommand{\otoprule}{\midrule[\heavyrulewidth]}

 \newcommand{\CH}[1]
    {CH-\MakeUppercase{\romannumeral #1}}   
 
\makeatletter
\renewcommand\env@matrix[1][\arraystretch]{%
  \edef\arraystretch{#1}%
  \hskip -\arraycolsep
  \let\@ifnextchar\new@ifnextchar
  \array{\c@MaxMatrixCols c}}
\makeatother

\newcommand{\by}{\boldsymbol{y}}
\newcommand{\bx}{\boldsymbol{x}}
\newcommand{\bX}{\boldsymbol{X}}
\newcommand{\bY}{\boldsymbol{Y}}

\newcommand{\bz}{\boldsymbol{z}}
\newcommand{\dv}{d_\mathsf{v}}
\newcommand{\dc}{d_\mathsf{c}}
\newcommand{\mL}{\mathcal{L}}
\newcommand{\Lf}{\mathsf{L}}
\newcommand{\mH}{\mathcal{H}}

\newcommand{\Tau}{\mathcal{T}}
\newcommand{\bs}[1]{\ensuremath{\boldsymbol{#1}}}    
\newcommand{\BP}{\text{BP}}
\newcommand{\MAP}{\text{MAP}}
\newcommand{\G}{\text{G}}
\newcommand{\hm}{\mathsf{h}}
\newcommand{\Eb}{E_{\text{b}}}
\newcolumntype{M}[1]{>{\centering\arraybackslash}p{#1}}

\makeatletter
\newcommand{\ostar}{\mathbin{\mathpalette\make@circled\star}}
\newcommand{\make@circled}[2]{%
  \ooalign{$\m@th#1\smallbigcirc{#1}$\cr\hidewidth$\m@th#1#2$\hidewidth\cr}%
}
\newcommand{\smallbigcirc}[1]{%
  \vcenter{\hbox{\scalebox{0.77778}{$\m@th#1\bigcirc$}}}%
}
\makeatother
     
\renewcommand{\boxed}[1]{\text{\fboxsep=.12em\fbox{$\displaystyle#1$}}}

\begin{document}
\title{Robust Performance Over Changing Intersymbol Interference Channels by Spatial Coupling } 

\author{

\IEEEauthorblockN{Mgeni Makambi Mashauri\IEEEauthorrefmark{1}, Alexandre Graell i Amat\IEEEauthorrefmark{2}, and Michael Lentmaier\IEEEauthorrefmark{1}}
%\vspace{0.05in}
\IEEEauthorblockA{\IEEEauthorrefmark{1}Department of Electrical and Information Technology, Lund University, Lund, Sweden}
\IEEEauthorblockA{\IEEEauthorrefmark{2}Department of Electrical Engineering, Chalmers University of Technology, Gothenburg, Sweden}\\\vspace*{-0.85cm}

\thanks{This work was supported in part by the Swedish Research Council (VR) under grant \#2017-04370. The simulations were performed on resources provided by the Swedish National Infrastructure for Computing (SNIC) at center for scientific and technical computing at Lund University (LUNARC).}
}
\maketitle

\begin{abstract}
We show that spatially coupled low-density parity-check~(LDPC) codes yield robust performance over changing intersymbol interfere~(ISI) channels with optimal and suboptimal detectors. We compare the performance with classical LDPC code design which involves optimizing the degree distribution for a given (known)  channel. We demonstrate that these classical schemes, despite working very good when designed for a given channel, can perform poorly if the channel is exchanged. %is unknown or even  changing.
With spatially coupled LDPC codes, however, we get performances close to the symmetric information rates with just a single code, without the need to know the channel and adapt to it at the transmitter. We also investigate threshold saturation with the linear minimum mean square error~(LMMSE) detector and show that with spatial coupling its performance can get remarkably close to that of an optimal detector for regular LDPC codes.     
\end{abstract}

\section{Introduction}
Spatially coupled codes were first studied in \cite{Fel99,Len10} in the context of low-density parity-check (LDPC) codes and later applied to other classes of codes\cite{Mol17}. They are known to exhibit remarkably good performance in a range of coding scenarios\cite{YNP2011,JBD2015,WHS2018} and scenarios beyond coding as well\cite{Kud2010,AVM2015}. This good performance is a consequence of the fact that the belief propagation~(BP) threshold of the coupled code approaches the threshold of the underlying uncoupled code with maximum a-priori~(MAP) decoding, a phenomenon known as threshold saturation. This phenomenon was proved in \cite{Kud2011,KRU2013} for binary memoryless symmetric~(BMS) channels.   
For such channels, it has been shown that spatially coupled LDPC codes can universally achieve capacity with BP decoding \cite{KRU2013}.

In \cite{Ngu2012}, it was shown that threshold saturation also occurs for channels with memory. The authors also showed that, with regular codes of high node degree, the BP threshold of the coupled code approaches the symmetric information rate~(SIR) of the simple dicode channel. In \cite{MIL2021}, the same phenomenon was demonstrated for three different intersymbol interference (ISI) channels with larger memory, and it was shown  that a single code  universally achieves the SIR of the three considered ISI channels. It can be observed that spatial coupling opens up a new paradigm of code design whereby the global MAP threshold, which was considered practically unattainable, now matters and can be achieved with the locally optimal BP decoding. One may then choose a code with good MAP threshold (which often implies bad BP threshold) and apply spatial coupling to attain the MAP threshold with BP decoding.
%%%%%%%%%%%%%%%%%%%%%%%%%%%%%%%%%%%%%%%%%%%%%%%%%%%%%%%%%%
\begin{figure}[t!]
\centerline{\includegraphics[scale=0.36]{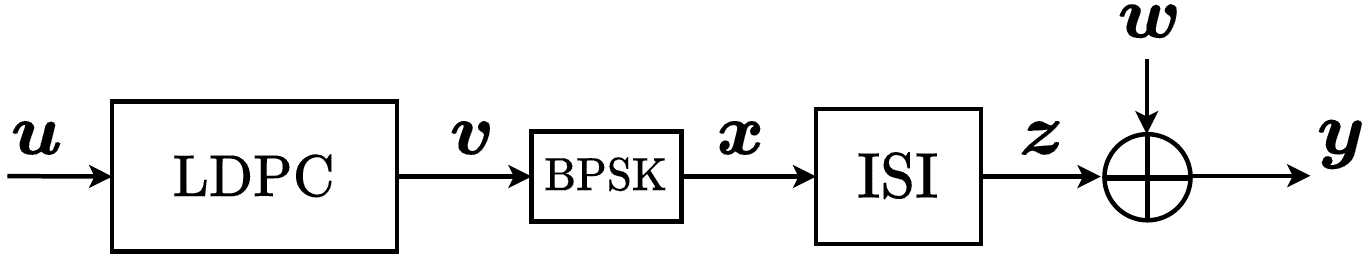}}
\caption{Block diagram showing the transmitter and the ISI channel. }
\label{transmitter}
\vspace{-2.5ex}
\end{figure}
%%%%%%%%%%%%%%%%%%%%%%%%%%%%%%%%%%%%%%%%%%%%%%%%%%%%%%%%%%

In this paper, we apply spatially coupled LDPC codes to turbo equalization demonstrating that their universality provides practical advantages when compared to classical code design. Classical code design for turbo equalization usually involves optimizing a code for a particular channel. For LDPC codes, this often means optimizing the degree distribution for the considered  channel \cite{TBr2004,KAV2003,KHG2012}. The shortcoming is that a degree distribution optimized for a given channel may perform poorly over a different channel. Furthermore,  one needs to know the channel at the transmitter. In contrast, due to their universal behavior, no optimization is required for spatially coupled LDPC codes, hence there is no need to know the channel and  they are expected to provide superior performance for scenarios where the channel changes. 
We compare the robustness of regular spatially coupled LDPC codes across three different ISI channels against optimized irregular LDPC codes in terms of  thresholds and finite length simulations. 
%The results we show, however, are for three different channels and not for  time varying channels since it is not even clear how one can effectively design an irregular code for random time varying channels. 
We also show that with regular codes the suboptimal linear minimum mean square error (LMMSE) detector, which has a very poor performance when compared to the optimal detector, has a performnce quite close to the BCJR detector when spatially coupled LDPC codes are used with both of them. 

The remainder of the paper is organized as follows. After introducing the considered system model in Section~\ref{sec:model}, we first discuss classical code design for ISI channels with irregular LDPC codes to highlight the weakness of such approach in Section~\ref{irregularDesign}. In Section~\ref{SpatailDesign}, we describe code design with spatially coupled LDPC codes for an optimal detector and how their universality can overcome these problems. In Section~\ref{SpatialMMSE}, we consider a suboptimal LMMSE  detector, discuss  threshold saturation  in this setting and 
show that we cannot simply use the area bound to provide a meaningful upper bound for the coupled threshold for such a detector. We then propose a method to approximate such threshold. Finally, the paper is concluded in Section~\ref{sec:Conclusion}.

\section{System Model}\label{sec:model}
The system model is shown in Fig.~\ref{transmitter}. A sequence of $k$ information bits $\boldsymbol{u}$ is encoded onto a codeword $\boldsymbol{v}$ of length $n$. The codeword is then mapped into a sequence of symbols $\boldsymbol{x}$ using binary phase shift keying~(BPSK) modulation with the mapping $0\mapsto +1$ and $1\mapsto -1$. The sequence $\boldsymbol{x}$ is transmitted over an ISI channel of memory $\nu$. The output of the channel filter, $\boldsymbol{z}$, is the convolution of $\boldsymbol{x}$ and the impulse response of the channel, $\boldsymbol{h}$, which has $\nu+1$ taps. Table~\ref{channels} shows the impulse responses of the considered ISI channels. The impulse responses are normalized such that  $||\bs{h}||=1$. This makes the signal energy at the receiver, $E_z$,  equal to the signal energy, $E_x$, at the input of the ISI channel. The received sequence, $\by$, is the result of corrupting $\bz$ by additive white Gaussian noise. We define $\gamma=\Eb/N_0$, where $\Eb$ is the average energy per information bit and  $N_0=2\sigma^2$  the noise spectrum density. Note that $E_{\text{b}}/N_0=E_z/RN_0$, where $R$ is the rate of the code.

\begin{figure}[t!]
\centerline{\includegraphics[scale=0.22]{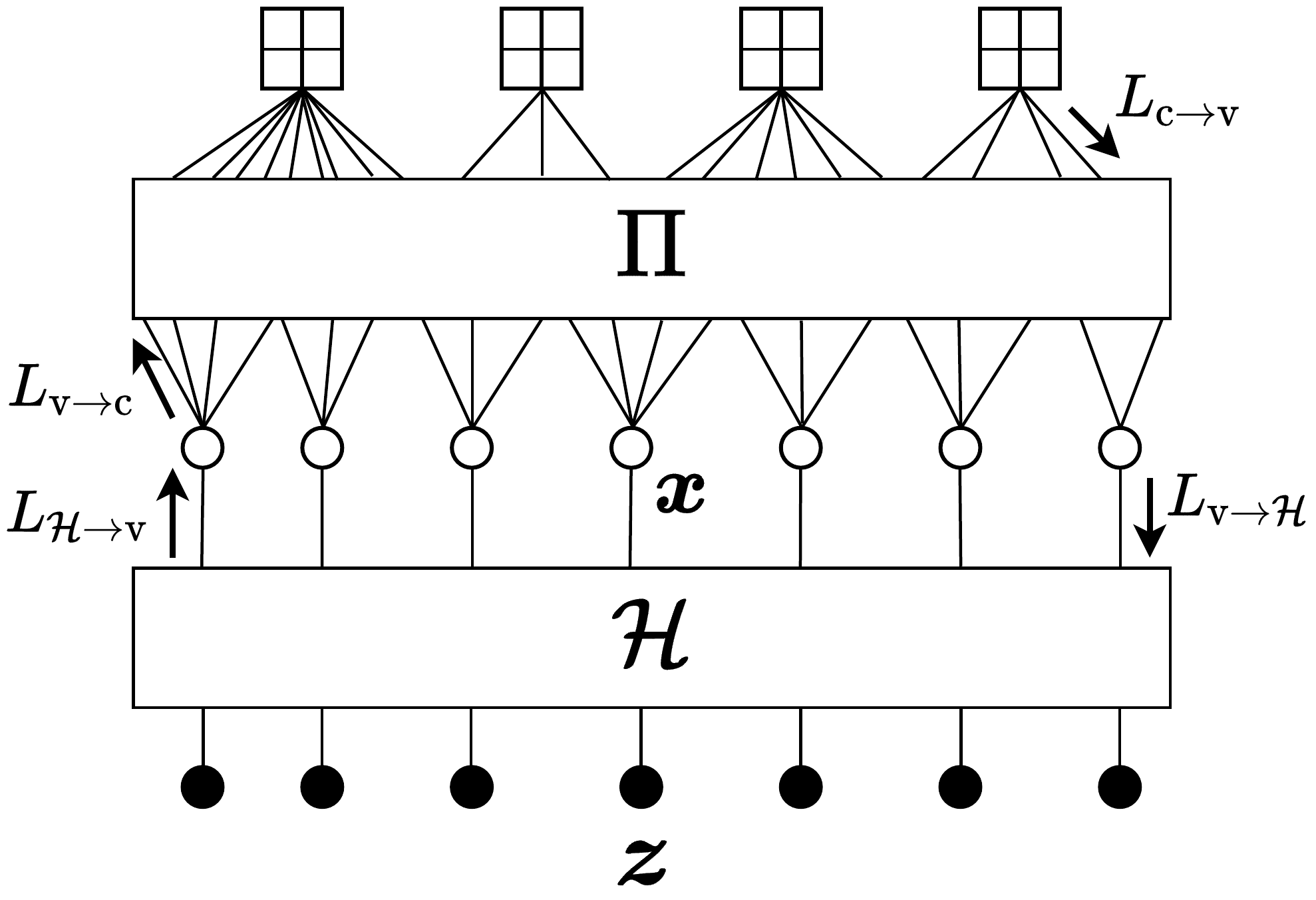}}
\vspace{-1.6ex}
\caption{Factor graph of turbo equalization with an irregular LDPC code. }
\label{irregular}
%\vspace{-2.0ex}
\end{figure}

At the receiver, the channel detector and the decoder exchange information iteratively, a process widely known as turbo equalization. We use BP decoding to decode the LDPC code while for the channel we consider two types of detectors\textemdash the optimal detector, implemented by applying the BCJR algorithm, and the suboptimal LMMSE detector. The detectors are implemented as described  in \cite{Tuch2011}. We thus have message exchanges within the decoder, i.e., between check nodes~(CNs) and variable nodes~(VNs) for $I_\text{C}$ iterations, and message exchanges between the VNs and the detector for $I_\text{D}$ iterations, as depicted by Fig.~\ref{irregular}. The messages exchanged are log-likelihood ratios. 
 %%%%%%%%%%%%%%%%%%%%%%%%%%%%%%%%%%%%%%%%%%%%%%%%%%%%%%%%%%
%%%%%%%%%%%%%%%%%%%%%%%%%%%%%%%%%%%%%%%%%%%%%%%%%%%%%%%%%%

\section{Code Design for Turbo Equalization with Irregular LDPC Codes}\label{irregularDesign}
We consider irregular LDPC codes with polynomials $\lambda(x)$ and $\Lf(x)$ representing the VN degree distribution  from an edge and node perspective, respectively, while the CN degree distribution from an edge perspective is represented by $\rho(x)$. The design rate for such a code is given by $R(\lambda,\rho)=1-\frac{\int_0^1 \rho(x)}{\int_{0}^1\lambda(x)}$. We consider a design rate of $1/2$ for all scenarios. 
To achieve good performance with turbo equalization with classical code design, the LDPC code needs to be specifically designed for a given channel. One approach is to optimize the degree distribution such that the BP threshold is maximized. This can be achieved via  density evolution. We briefly describe density evolution for irregular LDPC codes over an ISI channel.

The density passed from the detector to the code, $p(L^{(\ell)}_{\mH \rightarrow
\text{v}})$, is a function of the incoming densities from  all VNs  $p(L^{(\ell-1)}_{ \text{v} \rightarrow \mH})$ and the noise distribution. Denoting this function as $\Tau$ we have
\begin{align*}
p(L^{(\ell)}_{\mH \rightarrow \text{v}})=\Tau \left(p(L^{(\ell-1)}_{ \text{v} \rightarrow \mH}),\sigma \right)\,.
\end{align*}

It is not possible to obtain $\Tau(.,.)$ in closed form for Gaussian noise (it can be computed for erasure noise \cite{MIL2021}), but it can  be evaluated via Monte Carlo methods. For each VN, its outgoing density to an edge is the convolution of the density $p(L^{(\ell)}_{\mH \rightarrow \text{v}})$ from the detector and the $\dv-1$ incoming densities  from other edges, where $\dv$ is the VN degree. The average density from the VNs to a neighboring CN, $p(L^{(i)}_{\text{v} \rightarrow \text{c}})$,  is then obtained by averaging over the degree distribution $\lambda(x)$. At each CN of degree $\dc$, the outgoing density is computed from the $\dc-1$ incoming densities in a nested fashion using a two-dimensional lookup table for discretized density evolution \cite{Sae2001}. Similar to the VNs, the average density to a VN, $p(L^{(i)}_{\text{c} \rightarrow \text{v}})$, is obtained by averaging over the degree distribution $\rho(x)$. After $I_\text{C}$ iterations within the code, the density passed from a VN to the detector, $p(L^{(\ell)}_{ \text{v} \rightarrow \mH})$, is the convolution of the incoming $\dv$ densities from the neighboring CNs. The density evolution update equation for the joint BP decoding of the code and channel is thus given as
\begin{align*}
      p(L^{(i)}_{\text{v} \rightarrow \text{c}})&=p(L^{(\ell-1)}_{\mH \rightarrow \text{v}})\ostar\boldsymbol{\lambda}\left(p(L^{(i-1)}_{\text{c} \rightarrow \text{v}})\right)\\
      p(L^{(\ell)}_{\text{c} \rightarrow \text{v}})&=\boldsymbol{\rho}\left(p(L^{(i-1)}_{\text{v} \rightarrow \text{c}})\right)\\
    p(L^{(\ell)}_{ \text{v} \rightarrow \mH})&=\boldsymbol{\Lf}\left(L^{(I_\text{C})}_{\text{c} \rightarrow \text{v}})\right)  \\ p(L^{(\ell)}_{\mH \rightarrow \text{v}})&=\boldsymbol{\Tau}\left(p(L^{(\ell)}_{ \text{v} \rightarrow \mH})\right)
\end{align*}
 %%%
 where for a given density $a$, $\boldsymbol{\lambda}(a)=\sum_j\lambda_j a^{\ostar(j-1)}$, $\boldsymbol{\rho}(a)=\sum_j\rho_j a^{\boxed{*}(j-1)}$ and $\boldsymbol{\Lf}(a)=\sum_i\Lf_i a^{\ostar(j)}$. The operator $\ostar$ represents the convolution of densities while $\boxed{*}$ represents the density transformations at the CN as used in \cite{Ric2008}.
%%%%%%%%%%%%%%%%%%%%%%%%%%%%%%%%%%%%%%%%%%%%%%%%%%%%%%%%%%
\begin{table}[t]
\caption{Discrete impulse responses of the considered  ISI channels}
\vspace{-3.5ex}
\begin{center}
\begin{tabular}{cll}
	\toprule
	\CH{1} & $\bs{h}=[\begin{array}{cc} 0.7071 & -0.7071 \end{array}]$ &  $\nu=1 $\\[0.1cm]
	\CH{2} & $\bs{h}= [\begin{array}{ccc} 0.408 & 0.816&0.408 \end{array}]$ &  $\nu=2$\\[0.1cm]
	\CH{3} & $\bs{h}=[\begin{array}{ccccc} 0.227 & 0.46&0.688&0.46&0.227 \end{array}]$ &  $\nu=4 $\\[0.05cm]
	 \hline
\end{tabular}
\end{center}
\label{channels}
%\vspace{-5ex}
\end{table}
%%%%%%%%%%%%%%%%%%%%%%%%%%%%%%%%%%%%%%%%%%%%%%%%%%%%%%%%%% 

To find a code for a particular channel, we use a two-step searching scheme. In the first step,  a list of codes is generated using the EXIT chart design method. This is done by combining the code's VNs with the detector and fitting the EXIT curve of the combined node with that of the CNs. The method is used according to the description in \cite{TBr2004} but with the modification that we use CNs with more than one degree. The fitting is done manually at an SNR close to the SIR of the channel. The SIR values were calculated using a numerical method described in \cite{Arn2006}. Due to the approximate nature of the EXIT chart approach, a list of $N_\text{Z}$ codes whose curves closely fit (imperfections are allowed where we might have a small crossing of the curves) around the SIR rate are generated. In the second step, we perform density evolution for each of the $N_\text{Z}$ codes and select the code with the best BP threshold. Table~\ref{OptimizedBCJR} shows the codes obtained for the BCJR detector while those for the LMMSE detector are shown in Table~\ref{OptimizedLMMSE}.
With this procedure, we can find codes with threshold  close to the SIR of the  particular channel the code is designed for.   This approach, however, does not  guarantee that a code designed for a specific channel performs well for other ISI channels or a time-varying channel. 
Tables~\ref{changingBCJR} and \ref{changingLMMSE} show the BP thresholds of the designed codes for each of the three considered channels\footnote{With a slight abuse of notation, we use the term {\em BP thresholds} for the system with LMMSE detector as well, even though it is not locally optimal as the BCJR detector.}. In the tables each bold entry represents the threshold of a code matched to the channel it was designed for.
We observe that when a code designed for a given ISI channel is applied to a different  channel, the gap to the SIR can be  large. For example, the  code designed for \CH{2} under the LMMSE detector has a threshold $0.1$~dB  away from the corresponding SIR, but the code threshold of the same code is  nearly $3$~dB away for channel \CH{3}. 

%%%%%%%%%%%%%%%%%%%%%%%%%%%%%%%%%%%%%%%%%%%%%%%%%%%%%%%%%%
\begin{table}[!t]
\scriptsize
\caption{Codes optimized for the BCJR detector}
\vspace{-3.5ex}
\begin{center}
\begin{tabular}{M{0.15cm}M{0.65cm}M{0.65cm}|M{0.15cm}M{0.65cm}M{0.65cm}|M{0.15cm}M{0.65cm}M{0.65cm}}
\toprule
\multicolumn{3}{c|}{\CH{1}}    & \multicolumn{3}{c|}{\CH{2}}    & \multicolumn{3}{c}{\CH{3}}\\
\hline
$i$&$\lambda_i$& $\rho_i$  & $i$&$\lambda_i$& $\rho_i$  & $i$&$\lambda_i$& $\rho_i$ \\  
\otoprule
$2$ & $0.3075$  &             & $2$ & $0.3963$ &            & $2$ & $0.5935$ &           \\
$3$ & $0.3208$  &             & $3$ & $0.3589$ &            & $3$ & $0.0243$ &           \\
$5$ & $0.0180$  & $0.0436$    & $4$ &          & $0.0458$   & $6$ &          & $0.8856$  \\
$6$ & $0.0377$  &             & $5$ & $0.0153$ & $0.0189$   & $7$ &          & $0.1144$  \\
$7$ &           & $0.9456$    & $6$ &          & $0.9128$   & $11$& $0.0182$ &           \\
$8$ &           & $0.0108$    & $8$ &          & $0.0225$   & $17$& $0.3639$ &           \\
$10$& $0.0130$ &              & $9$ & $0.0592$ &            &     &          &           \\
$13$& $0.0446$ &              & $13$& $0.1583$ &            &     &          &           \\
$16$& $0.0876$ &              & $17$& $0.0120$ &            &     &          &           \\
$18$& $0.1708$ &              &     &          &            &     &          &           \\
\bottomrule                      
\end{tabular} 
\end{center}
\label{OptimizedBCJR}
\vspace{-2.8ex}
\end{table}
%%%%%%%%%%%%%%%%%%%%%%%%%%%%%%%%%%%%%%%%%%%%%%%%%%%%%%%%%%
%%%%%%%%%%%%%%%%%%%%%%%%%%%%%%%%%%%%%%%%%%%%%%%%%%%%%%%%%%
\begin{table}[!t]
\scriptsize
\caption{Codes optimized for the LMMSE detector}
\vspace{-3.5ex}
\begin{center}
\begin{tabular}{M{0.15cm}M{0.65cm}M{0.65cm}|M{0.15cm}M{0.65cm}M{0.65cm}|M{0.15cm}M{0.65cm}M{0.65cm}}
\toprule
\multicolumn{3}{c|}{\CH{1}}    & \multicolumn{3}{c|}{\CH{2}}    & \multicolumn{3}{c}{\CH{3}}\\
\hline
$i$&$\lambda_i$& $\rho_i$  & $i$&$\lambda_i$& $\rho_i$  & $i$&$\lambda_i$& $\rho_i$ \\  
\otoprule
$2$ & $0.2652$  &             & $2$  & $0.3131$  &            & $2$ & $0.4792$ & $0.0270$ \\
$3$ & $0.2921$  &             & $3$  & $0.2805$  &            & $3$ & $0.0357$ &           \\
$4$ & $0.0489$  & $0.0270$    & $4$  & $0.0123$  & $0.0496$   & $4$ & $0.0172$ & $0.0696$  \\
$5$ &           & $0.0663$    & $6$  &           & $0.1662$   & $5$ & $0.0209$ & $0.0150$  \\
$6$ & $0.0178$  &             & $8$  &           & $0.7841$   & $7$&           & $0.0632$  \\
$8$ &           & $0.9067$    & $10$ & $0.0139$  &            & $8$&           & $0.8252$  \\
$12$& $0.0306$  &             & $16$ & $0.2225$  &            & $19$& $0.4348$ &           \\
$14$& $0.0681$  &             & $20$ & $0.1578$  &            & $50$& $0.0122$ &           \\
$18$& $0.0419$  &             &      &           &            &     &          &           \\
$20$& $0.2355$  &             &      &           &            &     &          &           \\
\bottomrule                      
\end{tabular} 
\end{center}
\label{OptimizedLMMSE}
\vspace{-3.5ex}
\end{table}
%%%%%%%%%%%%%%%%%%%%%%%%%%%%%%%%%%%%%%%%%%%%%%%%%%%%%%%%%%

\section{Code Design with Spatially Coupled Codes}\label{SpatailDesign}
\begin{figure}[!t]
\centerline{\includegraphics[scale=0.20]{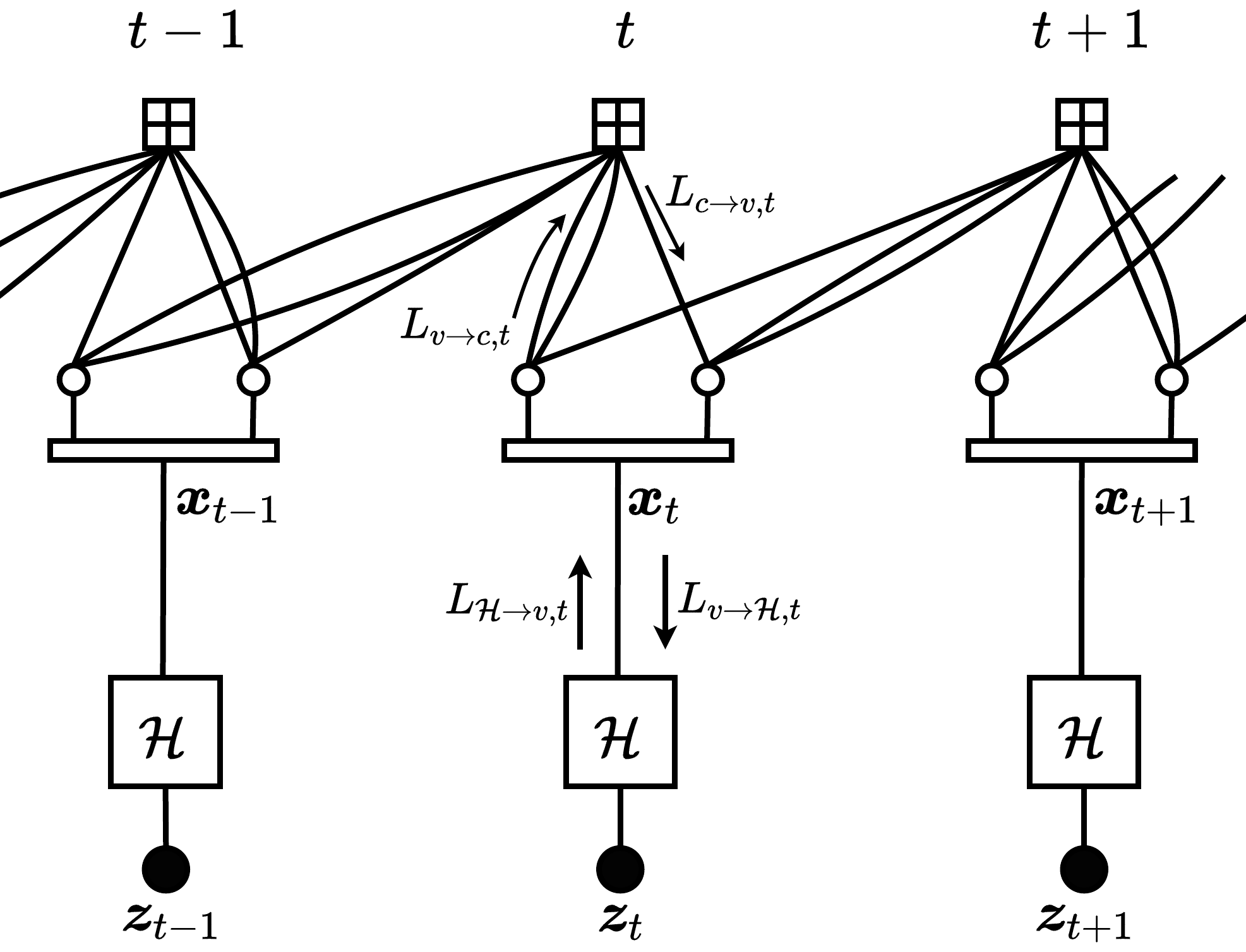}}
\vspace{-2ex}
\caption{Compact graph representation for equalization with a (3,6) SC-LDPC code with coupling memory $m=1$.}
\label{messages1}
\vspace{-2.5ex}
\end{figure}

We now consider spatially coupled LDPC codes with a BCJR detector. The whole system can be described by a factor graph which combines the graph representing the channel constraints and the Tanner graph of the LDPC code.  Specifically, the factor graph is constructed by placing $L$ copies of a $(\dv,\dc)$ regular LDPC code of VN degree $\dv$ and CN degree $\dc$ in $L$ spatial positions in the range $\mL\in\{1,\ldots,L\}$.  Fig.~\ref{messages1} shows the factor graph for three spatial positions, $t-1$, $t$, and $t+1$.  Each spatial position consists of $N$ VNs, represented by empty circles, and $M$ CNs $(M=\frac{\dv}{\dc}N)$, represented by squares with a cross. The $L$ copies are coupled as follows: each VN at position $t\in\mL$ is connected to CNs in the range $[t,\ldots,t+m]$, where $m$ is referred to as the coupling memory. Hence, each CN at position $t$ is connected to VNs in the range $[t-m,\ldots,t]$. 
The constraints of the ISI channel at each spatial position are  represented by a square labeled with the letter $\mH$, referred to as factor node.
Each of the VNs represented by the black circles at the bottom of the figure (denoted by $\{\bz_t\}$) represent a block of $N$ symbols at the output of the ISI channel before being corrupted by noise. This means that  $\bz=(\bz_1,\ldots,\bz_L)$ (see Fig.~\ref{transmitter}). The rectangles at each spatial position between the Tanner graph of the SC-LDPC code and the channel factor nodes represent multiplexers that multiplex the $N$ code bits  at each spatial position into a single binary sequence ($\bx_t$, with $\bx=(\bx_1,\ldots,\bx_L)$) at the input of the channel. This makes $n=NL$. 
Decoding is then performed by iteratively passing messages along the edges of the graph in Fig.~\ref{messages1}.
 
Owing to the universality of spatially coupled LDPC codes, we perform code design for ISI channels in two steps. We first pick a regular code with high node degree and then apply spatial coupling to achieve good performance over all channels.
 %%%%%%%%%%%%%%%%%%%%%%%%%%%%%%%%%%%%%%%%%%%%%%%
 \begin{table}[!t]
\caption{Code design and changing channel with BCJR}
\vspace{-2ex}
\begin{center}
\begin{tabular}{c|M{0.77cm}|M{0.77cm}|M{0.77cm}|M{0.77cm}}
\toprule
\multirow{2}*{Designed for }&\multicolumn{3}{c|}{BP threshold when applied to} & SIR\\
                           
                           &\CH{1}{}&\CH{2}&\CH{3}&{}\\
 \otoprule
 \CH{1} &$\mathbf{0.93}$&$1.65$&$4.55$&$0.82$\\[0.5mm]
 
  \CH{2} &$1.42$&$\mathbf{1.51}$&$4.11$&$1.44$\\[0.5mm]
 
  \CH{3} &$3.29$&$3.32$&$\mathbf{3.25}$&$2.96$ \\[0.5mm]

\bottomrule
\end{tabular}
\end{center}
\label{changingBCJR}
\vspace{-3ex}
\end{table}
 %%%%%%%%%%%%%%%%%%%%%%%%%%%%%%%%%%%%%%%%%%%%%%%%%%%%%%%
 %%
 %%
 %%%%%%%%%%%%%%%%%%%%%%%%%%%%%%%%%%%%%%%%%%%%%%%
 \begin{table}[!t]
\caption{Code design and changing channel with LMMSE}
\vspace{-2ex}
\begin{center}
\begin{tabular}{c|M{0.77cm}|M{0.77cm}|M{0.77cm}|M{0.77cm}}
\toprule
\multirow{2}*{Designed for }&\multicolumn{3}{c|}{BP threshold when applied to} & SIR\\
                           
                           &\CH{1}{}&\CH{2}&\CH{3}&{}\\
 \otoprule
 \CH{1} &$\mathbf{1.02}$&$1.96$&$6.53$&$0.82$\\[0.5mm]
 
  \CH{2} &$1.37$&$\mathbf{1.54}$&$5.80$&$1.44$\\[0.5mm]
 
  \CH{3} &$3.52$&$3.44$&$\mathbf{3.35}$&$2.96$ \\[0.5mm]
  
\bottomrule
\end{tabular}
\end{center}
\label{changingLMMSE}
\vspace{-5ex}
\end{table}
 %%%%%%%%%%%%%%%%%%%%%%%%%%%%%%%%%%%%%%%%%%%%%%%%%%%%%%%
 We apply this principle and consider the $(5,10)$ and $(6,12)$ regular LDPC codes. In Table~\ref{SpatialCombined} we give the thresholds for the uncoupled codes~(labeled $\gamma^{\BP}$) and those of the coupled codes for different coupling memories~(labelled $\gamma^{(m)}$). We also show the corresponding MAP thresholds and the SIRs as well.  
It is observed that  uncoupled regular LDPC codes  have  poor BP threshold, with a gap up to $5$~dB from the SIR. The MAP thresholds, however, are almost equal to the SIRs\footnote{The MAP thresholds for the system with BCJR detector were obtained from \cite{MIL2021} where they were obtained using the area theorem.}. Due to threshold saturation, the BP threshold of spatially coupled LDPC codes are close to the SIR for the three different ISI channels. For the $(6,12)$ code,  the threshold for memory $m=6$ is good for all channels and  better than the ones of the optimized codes in Table~\ref{changingBCJR}.

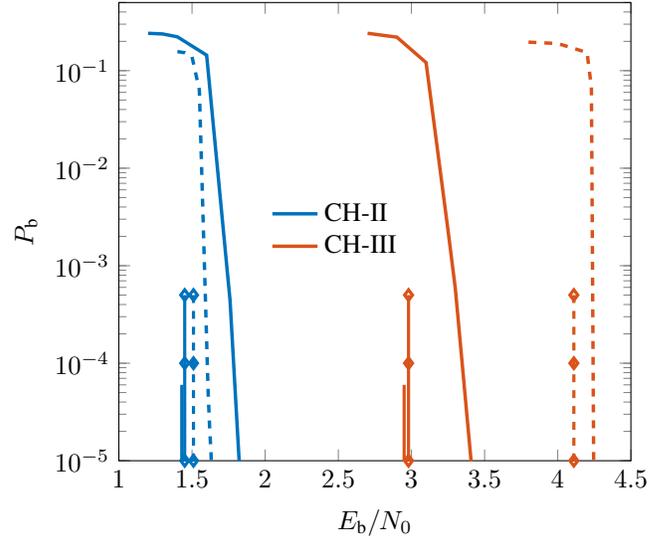
\begin{figure}[t]
\centering
% This file was created by matlab2tikz.
%
%The latest updates can be retrieved from
%  http://www.mathworks.com/matlabcentral/fileexchange/22022-matlab2tikz-matlab2tikz
%where you can also make suggestions and rate matlab2tikz.
%
\definecolor{mycolor1}{rgb}{0.00000,0.44700,0.74100}%
\definecolor{mycolor2}{rgb}{0.85000,0.32500,0.09800}%
\definecolor{mycolor3}{rgb}{0.9999,0.000,0.9999}%
\definecolor{mycolor4}{rgb}{0.80000,0.5100,0.70000}%
\definecolor{mycolor5}{rgb}{1.00000,0.00000,1.00000}%
\begin{tikzpicture}

\begin{axis}[%
width=2.68in,
height=2.4in,
at={(0.67in,3.5in)},
scale only axis,
xmin=1,
xmax=4.5,
xlabel={$\Eb/N_0$},
ymode=log,
ymin=1e-05,
ymax=0.5,
ylabel={$P_\text{b}$},
yminorticks=true,
axis background/.style={fill=white},
legend style={at={(0.28,0.42)}, anchor=south west, legend cell align=left, align=left, draw=white!15!white,fill=white},
]
\begin{comment}
\addplot [color=mycolor3, line width=1.0pt]
  table[row sep=crcr]{%
1	0.101\\
1.20	0.095268\\
1.30	0.064616\\
1.35	0.035396\\
1.40	0.01993224\\
1.42	0.01514536\\
1.45	0.01052172\\
1.50	0.00452304\\
1.60	0.00099184\\
1.70	0.00017432\\
1.75	7.246e-05\\
1.80	4.1e-05\\
1.90	1.476e-05\\
2.10	4.58e-06\\
};
\addlegendentry{CH-I}
\end{comment}

\addplot [color=mycolor1,  line width=1.3pt]
  table[row sep=crcr]{%
1.2 0.24181\\
1.3 0.2381\\
1.4 0.2225\\
1.6	0.1442\\
%1.75 0.0023\\
1.76	4.4075E-4\\
1.85 1.8141E-6\\
};
\addlegendentry{CH-II}

\addplot [color=mycolor2, line width=1.3pt]
  table[row sep=crcr]{%
2.7 0.2419\\
2.9 0.2209\\
3.1	0.1211\\
3.25 0.0023\\
3.3	5.8075E-4\\
3.45 1.9941E-6\\
};
\addlegendentry{CH-III}

\addplot [color=mycolor1, line width=1.2pt, mark=diamond, mark options={solid, mycolor1}]
  table[row sep=crcr]{%
1.45	5e-4\\
1.45	1e-4\\
1.45	1E-5\\
1.45	5E-6\\
};

\addplot [color=mycolor1, dashed, line width=1.3pt]
  table[row sep=crcr]{%
1.4	0.157064\\
1.45	0.152804\\
1.50	0.138722\\
1.55	0.063626\\
1.61	5.108321e-05\\
1.62	2.25631e-05\\
1.63	1.1762e-05\\
1.65	1.4215e-06\\
};

\addplot [color=mycolor2, dashed, line width=1.3pt]
  table[row sep=crcr]{%
%3	0.219824\\
%3.25	0.214704\\
%3.5	0.208236\\
3.8	0.196244\\
4	0.18996\\
4.2	0.153723\\
4.23	0.06785\\
4.25	7.8e-07\\
};

\addplot [color=mycolor2, line width=1.2pt]
  table[row sep=crcr]{%
2.95	6e-5\\
2.95	1E-5\\
2.95	5E-6\\
};

\addplot [color=mycolor2, line width=1.2pt, mark=diamond, mark options={solid, mycolor2}]
  table[row sep=crcr]{%
2.98	5e-4\\
2.98	1e-4\\
2.98	1E-5\\
2.98	5E-6\\
};
\addplot [color=mycolor2, dashed, line width=1.2pt, mark=diamond, mark options={solid, mycolor2}]
  table[row sep=crcr]{%
4.11	5e-4\\
4.11	1e-4\\
4.11	1E-5\\
4.11	5E-6\\
};

\addplot [color=mycolor1, dashed, line width=1.2pt, mark=diamond, mark options={solid, mycolor1}]
  table[row sep=crcr]{%
1.51	5e-4\\
1.51	1e-4\\
1.51	1E-5\\
1.51	5E-6\\
};
\addplot [color=mycolor1, line width=1.2pt]
  table[row sep=crcr]{%
1.43	6e-5\\
1.43	1E-5\\
1.43	5E-6\\
};

\end{axis}

\begin{axis}[%
width=5.833in,
height=4.375in,
at={(0in,0in)},
scale only axis,
xmin=0,
xmax=1,
ymin=0,
ymax=1,
axis line style={draw=none},
ticks=none,
axis x line*=bottom,
axis y line*=left,
legend style={legend cell align=left, align=left, draw=white!15!black}
]
\end{axis}
\end{tikzpicture}%
\vspace{-8.5cm}
\caption{Simulation results with BCJR detector. Dashed lines show the performance of the irregular code optimized for \CH{2} while solid lines are for that of the (6,12) with $m=6$. The solid short vertical lines are the corresponding SIRs while the long vertical lines with diamonds mark the  BP thresholds.}
\label{SimulationBCJR}
\vspace{-0.25cm}
\end{figure}

 In Fig.~\ref{SimulationBCJR}, we give simulation results  for the $(6,12)$ code for 
\CH{2} and \CH{3} and compare it to the irregular LDPC code designed for \CH{2}. For the spatially coupled LDPC code, we use $m=6$ and $L=500$ with $N=10 \, 000$, which gives the design rate $R=0.494$. The code is decoded using window decoding\cite{ZDL2017} with a window size of $W=30$, resulting in a decoding latency of $W N =300 \, 000$ symbols. Within the window, we use $I_\text{C}=30$ iterations in the code and $I_\text{D}=20$ iterations between the code and the channel. For the irregular code, we use a block length of $n=300 \, 000$ and the parity-check matrix is generated by the progressive edge growth algorithm \cite{XEA2005}. We use $I_\text{C}=30$ iterations within the code and $I_\text{D}=20$ iterations between the code and the channel. 
The results indicate a convergence to the thresholds in Tables~\ref{changingBCJR} and \ref{SpatialCombined}, whereby  the gaps are smaller for the uncoupled code. The irregular code designed for \CH{2} performs very well for that particular channel, but the performance deteriorates significantly when the channel is changed to \CH{3}. The coupled code shows good performance for both channels.      
%The results reflect the thresholds in Tables~\ref{changingBCJR} and \ref{SpatialCombined} whereby the irregular code designed for \CH{2} performs well for that channel but the performance deteriorate significantly when the channel is changed to \CH{3}. On the other hand the coupled code show good performance for both channels.      
%%%%%%%%%%%%%%%%%%%%%%%%%%%%%%%%%%%%%%%%%%%%%%%%%%%%%%%
%%%%%%%%%%%%%%%%%%%%%%%%%%%%%%%%%%%%%%%%%%%%%%%%%%%%%%%%%

 %%%%%%%%%%%%%%%%%%%%%%%%%%%%%%%%%%%%%%%%%%%%%%%%%%%%%%%%%%
 %%%%%%%%%%%%%%%%%%%%%%%%%%%%%%%%%%%%%%%%%%%%%%%%%%%%%%%
 
\section{Spatially Coupled Codes with the LMMSE Detector}\label{SpatialMMSE}

In this section, we consider the use of spatially-coupled LDPC codes with a  suboptimal channel detector, namely the LMMSE detector. For the LMMSE detector, we also observe in Table~\ref{SpatialCombined} that the BP threshold of spatially coupled LDPC codes improves with increasing memory. It is not clear, however, which value the coupled threshold saturates to. This is because using the generalized extrinsic information transfer~(GEXIT) bounding technique as for the BCJR detector will not give us a meaningful bound. This can be observed by noting that the upper bound to the MAP threshold is based on the following facts. The GEXIT of an ISI channel with entropy $\hm=\text{H}(Z|Y)$ and initial state $S_0$ is defined as\cite{Ngu2012} 
 $$
 \G(\hm)=\frac{1}{n} \frac{\text{d}\text{H}(\bX|\bY,S_0)}{\text{d}\hm} .
 $$
 
 The conditional entropy rate $\text{H}(\bX|\bY,S_0)$ can be thought of as the output of the globally optimal receiver~(MAP receiver). But if we use a locally optimal receiver (using BP) instead, the output entropy  will be greater than or equal to that of a globally optimal receiver. This is because the globally optimal receiver will reduce the uncertainty about $\bX$ compared to the locally optimal receiver. By definition, the integral of the GEXIT equals the rate of the code as $n$ grows to infinity and  it is equal to zero below the MAP threshold. We thus have 
 $$
R=\lim_{n \to \infty} \int_{\hm^\MAP}^{1}\G(\hm)\leq  \int_{\hm^\MAP}^{1}\G^{\BP}(\hm)\,,
 $$

 where $\G^{\BP}(\hm)$ is obtained from a BP decoder with both the block length $n$ and number of iterations $I_\text{D}$ approaching infinity.
 This provides a way to compute an upper bound for the MAP threshold\cite{MMU2009}. This is done by finding $\bar{\hm}$ such that it is the largest positive number such that
 $$
 \int_{\bar{\hm}}^{1}\G^{\BP}(\hm)=R .
 $$
 This bound is tight for a receiver with the BCJR detector. If we
change the detector to a suboptimal detector like the LMMSE, the message passing receiver is no longer locally optimal thus the performance of the whole sytem is degraded. As a result, the LMMSE detector will result into greater entropy than the BCJR detector. This implies that
 $$
  \int_{\hm^\MAP}^{1}\G^{\text{BCJR}}(\hm)\leq  \int_{\hm^\MAP}^{1}\G^{\text{LMMSE}}(\hm).
 $$
 Thus if we were to apply the same bounding technique, we would have an upper bound higher than that of the receiver with the BCJR detector. But since we know that the performance will be degraded by using a suboptimal detector, this bound becomes meaningless.

 We can however approximate this value using the positive gap condition for the EXIT curves\cite{KdUb2015}. This is done by combining the detector and the  VNs of the LDPC code and computing the area between the EXIT curve of the combined node, $h_\text{f}$, and that of the CN, $h_\text{g}$. An estimate of the BP threshold of the coupled system is given by finding the SNR at which the area between the curves transitions from negative to positive as $\Eb/N_0$ is increased.
\begin{table*}[!t]
%\scriptsize
\caption{Thresholds of regular codes with spatial coupling for ISI channels with BCJR and LMMSE detectors}
\vspace{-3.5ex}
\begin{center}
\begin{tabular}{cc|cccccc|cccccc}
\toprule
{}&{}&\multicolumn{6}{c|}{BCJR}&\multicolumn{6}{c}{LMMSE}\\

Code&Channel&$\gamma^{\BP}$&$\gamma^{(1)}$&$\gamma^{(6)}$&$\gamma^{(10)}$&$\gamma^{\MAP}$&$\gamma^{\text{SIR}}$&$\gamma^{\BP}$&$\gamma^{(1)}$&$\gamma^{(6)}$&$\gamma^{(10)}$&$\gamma^{\text{Area}}$&$\gamma^{\text{SIR}}$\\
\otoprule
\multirow{3}*{$(5,10)$}&\CH{1}&$3.03$ &$0.88$ &$0.85$ &$0.84$&$0.83$&$0.82$&  $3.55$& $0.98$& $0.97$& $0.97$& $1.08$& $0.82$\\[0.5mm]
                      &\CH{2}&$4.35$ &$1.483$&$1.45$  &$1.45$&$1.44$&$1.44$&    $5.84$& $1.71$& $1.68$& $1.68$& $1.82$& $1.44$\\[0.5mm]
                       &\CH{3}&$7.55$ &$3.04$ &$2.99$  &$2.99$&$2.96$&$2.96$&   $13.31$&$3.62$&$3.59$&$3.58$&$3.71$&$2.96$\\[0.5mm]
\hline
\multirow{3}*{$(6,12)$}&\CH{1}&$3.52$ &$0.85$ &$0.84$ &$0.83$&$0.82$&$0.82$&  $4.18$& $0.98$& $0.96$& $0.96$& $1.12$& $0.82$\\[0.5mm]
                      &\CH{2}&$4.94$ &$1.48$&$1.45$ &$1.45$&$1.44$&$1.44$&    $7.00$& $1.71$& $1.68$& $1.67$& $1.85$& $1.44$\\[0.5mm]
                       &\CH{3}&$8.12$ &$3.04$ &$2.98$&$2.98$&$2.96$&$2.96$&   $14.15$&$3.69$&$3.59$&$3.56$&$3.72$&$2.96$\\
\bottomrule                      
\end{tabular} 
\end{center}
\label{SpatialCombined}
\vspace{-1ex}
\end{table*}
%%%%%%%%%%%%%%%%%%%%%%%%%%%%%%%%%%%%%%%%%%%%%%%%%%%%%%%%%%
If we denote the function which gives the entropy of a symmetric Gaussian with mean $\mu$ by $\psi(\mu)$, the average output entropy from VNs  to a CN, $\text{h}_{\text{E,VN}}$, is given by 
\begin{equation}\label{hfromVNtoCN}
   \text{h}_{\text{E,VN}}=\sum_{i}\lambda_i\psi\Big((i-1)\psi^{-1}(\text{h}_{\text{E,CN}})+\psi^{-1}(\text{h}_{\text{E,DET}}) \Big)
\end{equation}
where $\text{h}_{\text{E,CN}}$ is the average entropy from a CN and $\text{h}_{\text{E,DET}}$ is the output entropy from a detector. The output entropy of the detector is a function of the a-priori entropy to the detector $\text{h}_{\text{A,DET}}$ and $E_z/N_0$,
$$
  \text{h}_{\text{E,DET}}=\mathcal{D}\left(\text{h}_{\text{A,DET}},\frac{E_z}{N_0}\right).  
$$
The function $\mathcal{D}$ is computed by Monte Carlo simulations and for a given $E_z/N_0$ it can be approximated by a third order polynomial in the a-priori entropy,
\begin{equation}\label{hfromchannel}
    \mathcal{D}\left(\text{h}_{\text{A,DET}},\frac{E_z}{N_0}\right)=c_3\text{h}_{\text{A,DET}}^3+c_2\text{h}_{\text{A,DET}}^2+c_1\text{h}_{\text{A,DET}}+c_0\hspace{0.07cm} , \
\end{equation}
%%%%%%%
where
%$\begin{array}{cccc} c_3 & c_2 & c_1 & c_0 \end{array}$
\begin{equation}\label{htochannel}
   \text{h}_{\text{A,DET}}=\sum_{i}L_i\psi\Big(i\psi^{-1}(\text{h}_{\text{E,CN}})\Big) .\
\end{equation}
Substituting \eqref{htochannel} and \eqref{hfromchannel} into \eqref{hfromVNtoCN}, we obtain the EXIT curve of the combined variable-detector node $h_\text{f}$ for a given SNR.

For the CN, we use the dual approximation \cite[p.~236]{Ric2008} to obtain the output entropy as
$$
\text{h}_{\text{E,CN}}=1-\sum_{i}\rho_i\psi \Big((i-1)\psi^{-1}(1-\text{h}_{\text{E,VN}})\Big)\,.
$$
We use a numerical approximation for $\psi(\mu)$ and $\psi^{-1}(\mu)$ based on the numerical approximations given in \cite{TBr2004} for the $J(\sigma)$ function which gives the mutual information of a symmetric Gaussian density. We use the fact that for a symmetric Gaussian density, $\mu=\frac{\sigma^2}{2}$ which implies $J(\sigma)=1-\psi(\frac{\sigma^2}{2})$. We thus have the recursion
\begin{align*}
    h_\text{g}(v)&=1-\sum_{i}\rho_i\psi \Big((i-1)\psi^{-1}(1-v)\Big)\\
    h_\text{f}(u)&=\sum_{i}\lambda_i\psi\Big((i-1)\psi^{-1}(u)+\psi^{-1}\big(\mathcal{D}_\text{L}(u)\big) \Big)
\end{align*}
where 
$$
\mathcal{D}_\text{L}(u)=\mathcal{D}\Big(\sum_{i}L_i\psi\big(i\psi^{-1}(u)\big),\frac{E_z}{N_0} \Big) .\
$$
%%%%%%%%%%%%%%%%%%%%%%%%%%%%%%%%%%%%%%%%%%%%%%%%%%%%%%%%%%

\begin{figure}[t]
\centering
\input{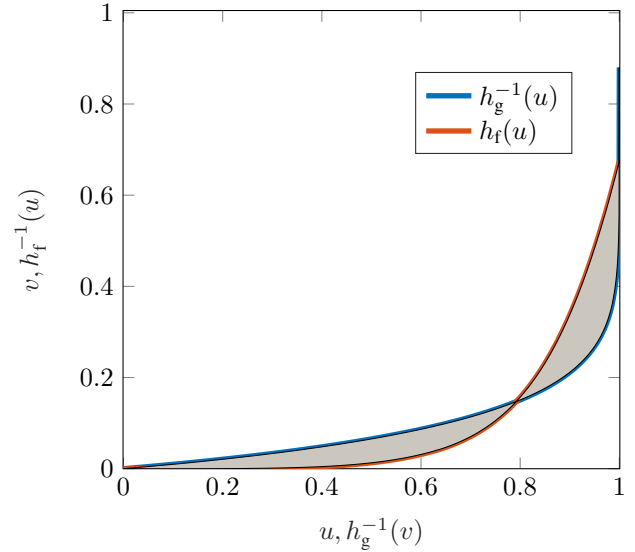}
\vspace{-8.5cm}
\caption{Approximating the coupled threshold of an (6,12) LDPC code and LMMSE equalizer for \CH{2} using the EXIT chart paradigm. The net area between the curve is zero at $\frac{\Eb}{N_0}=1.85$~dB. }
\label{potentialMMSE}
\vspace{-0.35cm}
\end{figure}

Fig.~\ref{potentialMMSE} illustrates how this recursion can be used to approximate the coupled threshold for the LMMSE detector. In Table~\ref{SpatialCombined}, we denote this approximate value by $\gamma^{\text{Area}}$. It can be seen that the coupled threshold for higher memory seems to saturate to a value close but not equal to $\gamma^{\text{Area}}$. It appears that $\gamma^{\text{Area}}$ is a pessimistic estimate as the coupled threshold always exceeds it. On the other hand, $\gamma^{\text{Area}}$ provides an efficient way to roughly predict the coupling gain directly from the uncoupled graph.  %But $\gamma^{\text{Area}}$ is still a quick way to predict the performance of the coupled system without implementing and simulating it.
%%%%%%%%%%%%%%%%%%%%%%%%%%%%%%%%%%%%%%%%%%%%%%%%%%%%
 %%%%%%%%%%%%%%%%%%%%%%%%%%%%%%%%%%%%%%%%%%%%%%%%%%%%%%%

We can notice in Table~\ref{SpatialCombined} that the LMMSE detector without coupling performs quite bad when compared to optimized irregular codes in Table~\ref{OptimizedLMMSE}. With coupling, however, the thresholds are improved significantly. In contrast to the case with the BCJR detector in Section~\ref{SpatailDesign}, the coupled codes with the LMMSE detector do not always outperform the optimized codes for the matched channels but they are nevertheless very close to them and we still have the robustness with changing channels.

\begin{figure}[t]
\centering
% This file was created by matlab2tikz.
%
%The latest updates can be retrieved from
%  http://www.mathworks.com/matlabcentral/fileexchange/22022-matlab2tikz-matlab2tikz
%where you can also make suggestions and rate matlab2tikz.
%
\definecolor{mycolor1}{rgb}{0.00000,0.44700,0.74100}%
\definecolor{mycolor2}{rgb}{0.99000,0.11500,0.09800}%
\definecolor{mycolor3}{rgb}{0.4500,0.95000,0.79800}%
\begin{tikzpicture}

\begin{axis}[%
width=2.75in,
height=2.56in,
at={(0.481in,0.346in)},
scale only axis,
xmin=2,
xmax=16,
xlabel={$\Eb/N_0$},
ymode=log,
ymin=1e-05,
ymax=0.5,
ylabel={$P_b$},
yminorticks=true,
axis background/.style={fill=white},
%xmajorgrids,
%ymajorgrids,
%yminorgrids,
legend style={at={(0.51,0.574)}, anchor=south west, legend cell align=left, align=left, draw=white!15!black}
]
\addplot [color=mycolor1,  line width=1.3pt]
  table[row sep=crcr]{%
2.7 0.2419\\
2.9 0.2209\\
3.1	0.1211\\
3.25 0.0023\\
3.3	5.8075E-4\\
3.45 1.9941E-6\\
};
\addlegendentry{BCJR}

\addplot [color=mycolor2, line width=1.3pt]
  table[row sep=crcr]{%
%3.0	0.16841\\
3.3	0.15741\\
3.4	0.15375\\
3.55	0.14625\\
3.7	0.1357\\
3.8	0.13275\\
3.9	0.10276\\
3.95	0.025159\\
4.05	6.2286e-05\\
4.1	1.2252e-06\\
};
\addlegendentry{LMMSE}

\addplot [color=black, line width=1.5pt]
  table[row sep=crcr]{%
2.96	1e-4\\
2.96	1E-5\\
2.96	5E-6\\
};
\addlegendentry{SIR}

\addplot [color=mycolor1, dashed, line width=1.3pt]
  table[row sep=crcr]{%
6	0.185068\\
7	0.1558\\
8	0.0097868\\
8.15	0.000928888888888889\\
8.2	1.0272e-05\\
8.21	0\\
};

\addplot [color=mycolor2, dashed, line width=1.3pt]
  table[row sep=crcr]{%
%4.5	0.226636\\
%7.5	0.196796\\
9.5	0.177124\\
12.5	0.139484\\
13.5	0.125184\\
14	0.114916\\
14.5	1.04332e-05\\
};

\begin{comment}
\addplot [color=mycolor3, line width=1.5pt]
  table[row sep=crcr]{%
2.65	0.222392\\
2.85	0.219154\\
3.1	0.21276\\
3.15	0.211064\\
3.2	0.208802\\
3.25	0.20532\\
3.35	0.202562\\
3.45	4.7719e-06\\
};
\addlegendentry{IREG}

\addplot [color=mycolor3, line width=1.5pt, mark=o, mark options={solid, mycolor3}]
  table[row sep=crcr]{%
3	0.227619047619048\\
3.5	0.221912380952381\\
4	0.211154285714286\\
4.5	0.19943619047619\\
4.65	0.00060124\\
4.75	2.8712e-06\\
};
\end{comment}

\end{axis}

\begin{axis}[%
width=3.702in,
height=3.148in,
at={(0in,0in)},
scale only axis,
xmin=0,
xmax=1,
ymin=0,
ymax=1,
axis line style={draw=none},
ticks=none,
axis x line*=bottom,
axis y line*=left,
legend style={legend cell align=left, align=left, draw=white!15!black}
]
\end{axis}
\end{tikzpicture}%
\vspace{-0.8cm}
\caption{Simulation results comparing the BCJR vs LMMSE detector without coupling~(dashed lines) and with coupling~(solid lines) for a (6,12) LDPC code and \CH{3}.}
\label{DetectorChange}
\vspace{-0.51cm}
\end{figure}
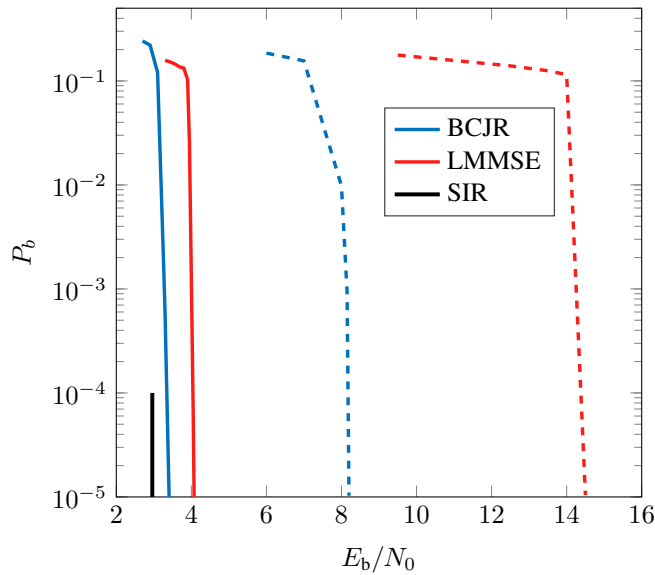

We can also notice in Table~\ref{SpatialCombined} that the linear detector without coupling yields poor thresholds when compared to the optimal detector for the same code and channel, especially for channels with severe ISI~(for example in \CH{2} and \CH{3}). The situation, however, changes significantly when spatial coupling is introduced. We see that the linear detector achieves performance very close to the optimal detector with spatial coupling. For \CH{3}, for example, the BP threshold of the $(5,10)$ and $(6,12)$ codes with LMMSE detector without coupling is more than 5~dB away from the threshold with the BJCR detector.
With coupling, however, the gap is only a fraction of dBs for all coupling memories. In Fig.~\ref{DetectorChange} this effect is shown with simulation results for the (6,12) code with the same settings as for the BCJR case discussed above for \CH{3}. In the figure we can clearly notice the narrowing of the gap between the two detectors when spatially coupled LDPC codes are applied.

\section{Conclusions}\label{sec:Conclusion}

We have demonstrated that spatially coupled LDPC codes are robust against both changing channel conditions and changing detector type when compared to classical code design. We can just use one LDPC code with high node degree with spatial coupling and attain universally good results for a number of ISI channels with BP decoding. It also makes the use of the suboptimal linear MMSE detector quite competitive with the BCJR detector. This can be of great practical value especially with situations in which applying the BCJR detector can be prohibitively complex due to, e.g., large channel memory.   
 %%
 %%%%%%%%%%%%%%%%%%%%%%%%%%%%%%%%%%%%%%%%%%%%%%%%%%%%%%
 %%%%%%%%%%%%%%%%%%%%%%%%%%%%%%%%%%%%%%%%%%%%%%%%%%%%%%%
\newpage
%\bibliographystyle{IEEEtran}
%\bibliography{references.bib} 
% Generated by IEEEtran.bst, version: 1.12 (2007/01/11)

\end{document}